\documentclass[referee, pdflatex, sn-bacis]{sn-jnl}% Basic Springer Nature Reference Style/Chemistry Reference Style

\jyear{2022}%
\usepackage{subfigure}
\theoremstyle{thmstyleone}%
\theoremstyle{thmstyletwo}

\theoremstyle{thmstylethree}

\unnumbered% uncomment this for unnumbered level heads

\begin{document}

\title[Multi-objective Generative Design of Three-Dimensional Composite Materials]{Multi-objective Generative Design of Three-Dimensional Composite Materials}


\author[1]{\fnm{Zhengyang} \sur{Zhang}}\email{zhengyangzhang@stju.edu.cn}
\equalcont

\author[1]{\fnm{Han} \sur{Fang}}\email{han.fang@sjtu.edu.cn}
\equalcont{These authors contributed equally to this work.}

\author[2]{\fnm{Zhao} \sur{Xu}}\email{xuzhao@csu.ac.cn}

\author[3]{\fnm{Jiajie} \sur{Lv}}\email{jiajie.lv@gmail.com}

\author[4]{\fnm{Yao} \sur{Shen}}\email{yaoshen@sjtu.edu.cn}

\author*[1]{\fnm{Yanming} \sur{Wang}}\email{yanming.wang@stju.edu.cn}

\affil[1]{\orgdiv{University of Michigan-Shanghai Jiao Tong University Joint Institute}, \orgname{Shanghai Jiao Tong University}, \orgaddress{\city{Shanghai}, \postcode{200240}, \country{China}}}
\affil[2]{\orgdiv{Key Laboratory of Space Utilization, Technology and Engineering Center for Space Utilization}, \orgname{Chinese Academy of Sciences}, \orgaddress{\city{Beijing}, \postcode{100094}, \country{China}}}
\affil[3]{\orgdiv{School of Materials Science and Engineering}, \orgname{Jilin University}, \orgaddress{\city{Changchun}, \postcode{130012}, \country{China}}}
\affil[4]{\orgdiv{School of Materials and Engineering}, \orgname{Shanghai Jiao Tong University}, \orgaddress{\city{Shanghai}, \postcode{200240}, \country{China}}}

\clearpage
\thispagestyle{empty}

 \abstract{Composite materials with 3D architectures are desirable in a variety of applications for the capability of tailoring their properties to meet multiple functional requirements. By the arrangement of materials’ internal components, structure design is of great significance in tuning the properties of the composites. However, most of the composite structures are proposed by empirical designs following existing patterns. Hindered by the complexity of 3D structures, it is hard to extract customized structures with multiple desired properties from large design space. Here we report a multi-objective driven Wasserstein generative adversarial network (MDWGAN) to implement inverse designs of 3D composite structures according to given geometrical, structural and mechanical requirements. Our framework consists a GAN based network which generates 3D composite structures possessing with similar geometrical and structural features to the target dataset. Besides, multiple objectives are introduced to our framework for the control of mechanical property and isotropy of the composites. Real time calculation of the properties in training iterations is achieved by an accurate surrogate model. We constructed a small and concise dataset to illustrate our framework. With multiple objectives combined by their weight, and the 3D-GAN act as a soft constraint, our framework is proved to be capable of tuning the properties of the generated composites in multiple aspects, while keeping the selected features of different kinds of structures. The feasibility on small dataset and potential scalability on objectives of other properties make our work a novel, effective approach to provide fast, experience free composite structure designs for various functional materials.
}

%%================================%%
%% Sample for structured abstract %%
%%================================%%

% \abstract{\textbf{Purpose:} 
% \textbf{Methods:} 
% \textbf{Results:} 
% \textbf{Conclusion:} }

\keywords{Composite Material; Composite Structures; Generative Design; 3D Wasserstein Generative Adversarial Network; Multi-Objective.}



\maketitle

\section{Introduction}\label{sec1}
% Composite maybe too large a topic here?%
Composites have been widely used in aerospace\cite{bib6_Rawal2001}, biomedicine\cite{bib7_Hong2015, bib32_Feig2018} and energy-related\cite{bib3_wang2018} applications for their advantages in mechanical\cite{bib12_Wegner2000, bib26_Frey2019}, thermal\cite{bib2_han2005, bib35_Zhou2021}, electronic\cite{bib3_wang2018}, impact resistance\cite{bib4_huang2020} and damage tolerance \cite{bib5_Li2018, bib14_Umanzor2021}properties. In recent decades, the development of architectured-materials\cite{bib8_Schaedler2011, bib9_Pham2019} and novel manufacturing techniques\cite{bib36_Chen2022} such as additive manufacturing\cite{bib10_Hong2015, bib11_Zhang2020} facilitates a precise control of architectures and micro-structures of  constituent phases in composite materials. Composite materials with various internal structures are manufactured to tune the overall performance for their functional requirements. Thus, to provide guidelines for composite structure design and manufacturing, the structure-property relationship of the composites have become a research hot spot. The mechanical properties of structures such as  regular lattices reinforced (including simple cubic, body-centered cubic, face-centered cubic and various auxetic structures)\cite{bib5_Li2018, bib13_Hu2019, bib14_Umanzor2021, bib15_Hu2019, bib16_Abdelhamid2018, bib17_Helou2018, bib18_Li2018}, stochastic fibre networks reinforced (i.e. random fibre networks and Voronoi fibre networks)\cite{bib11_Zhang2020, bib19_Zhang2013, bib20_Lin2019}, and triply periodic minimal surfaces reinforced composites (categorized as Spinodal Shell/Solid, Schwarz Primitive, etc.)\cite{bib22_Abueidda2015,bib21_Abueidda2016, bib23_Al-Ketan2017, bib24_Al-Ketan2017, bib25_Zhang2021} have been tuned by adjusting their structural designs.

However, the majority of studies focused on a specific type of composite structure by empirical or trial-and-error methods. A structural design space much larger than currently popular composite structures remains unexplored. It is still hindered by the structural diversity and complexity of 3D composite structures to explore the huge structural design space for proposing new structures.  Modern machine learning methods\cite{bib27_Cecen2018, bib28_Xue2020, bib29_Liu2015, bib30_Liu2017, bib31_Yang2018, bib33_Latypov2017} 
are considered as promising approaches to fulfill this kind of objectives and have been used to help the property prediction and design of various materials including polymer membranes\cite{bib47_Barnett2020}, alloys\cite{bib48_Wen2019, bib49_Roy2020}, and solid-state electrolytes\cite{bib50_Wang2020}. In terms of composites, machine learning models (mainly convolutional neural network, CNN) showed both good accuracy (95\%) and efficiency (250 times faster than finite element method, FEM) in predicting mechanical properties of 2D structures and pinpointed high performers for toughness or strength\cite{bib45_GU201819}.  Cecen et. al.\cite{bib29_Liu2015} and Liu et. al\cite{bib27_Cecen2018} built structure-property regression models for composites by M5 model tree and CNN algorithms.  Inputting 2D composite structures as images to machine learning models, thermal conductivity predictions of composites were also achieved via support vector regression (SVR), Gaussian process regression (GPR) and CNN\cite{bib46_WEI2018908}.

Among the machine learning approaches, generative models have been proved promising to assist the inverse design of compositions\cite{bib51_Rao2022}, molecular configurations\cite{bib38_Schutt2017, bib39_Schutt2017, bib40_NIPS2019_8974, bib37_Gebauer2022, bib52_Noh2019} as well as macro structures. For example, variational autoencoders (VAE) were introduced to provide optimal combinations of multiple representative structures of two-dimensional composite materials with the largest elastic properties\cite{bib28_Xue2020}. Mao et. al. utilized generative adversarial networks (GAN) to generate two-dimensional porous isotropic structures with various porosities and high Young's moduli\cite{bib34_Mao2020}. However, those works about structures are mostly based on 2D structures, or only generate similar structures with no other design objectives involved\cite{bib41_Kench2021}. The generative design under physics-based objectives of 3D composite structures still haven't been achieved. 

In this paper, we aim to develop a multi-objective driven, three-dimensional generative design framework for composite structures by integrating physics based controls into data driven methods. Wasserstein Generative Adversarial Network (WGAN) is extended to 3D space as a soft constraint to ensure the structural features similarity between generated structures and those in the training dataset, while 3D ResNet based surrogate model is trained to give real-time evaluations of their elastic characters including Young's moduli and isotropy. We introduce multiple design objectives including structural symmetry, elastic moduli, and mechanical isotropy fulfilled by mechanics based loss functions. We construct relatively small and concise dataset by automated finite element (FE) calculations of composite structures. Our generated structures demonstrate the capability of our framework to tuned performance for composite materials to meet given objectives while keeping targeted geometrical and structural features.

\section{Results}\label{sec2}

\subsection{3D MDWGAN Framework}\label{mdwgan}
\begin{figure}[htbp]
\centering
\includegraphics[width=0.9\linewidth]{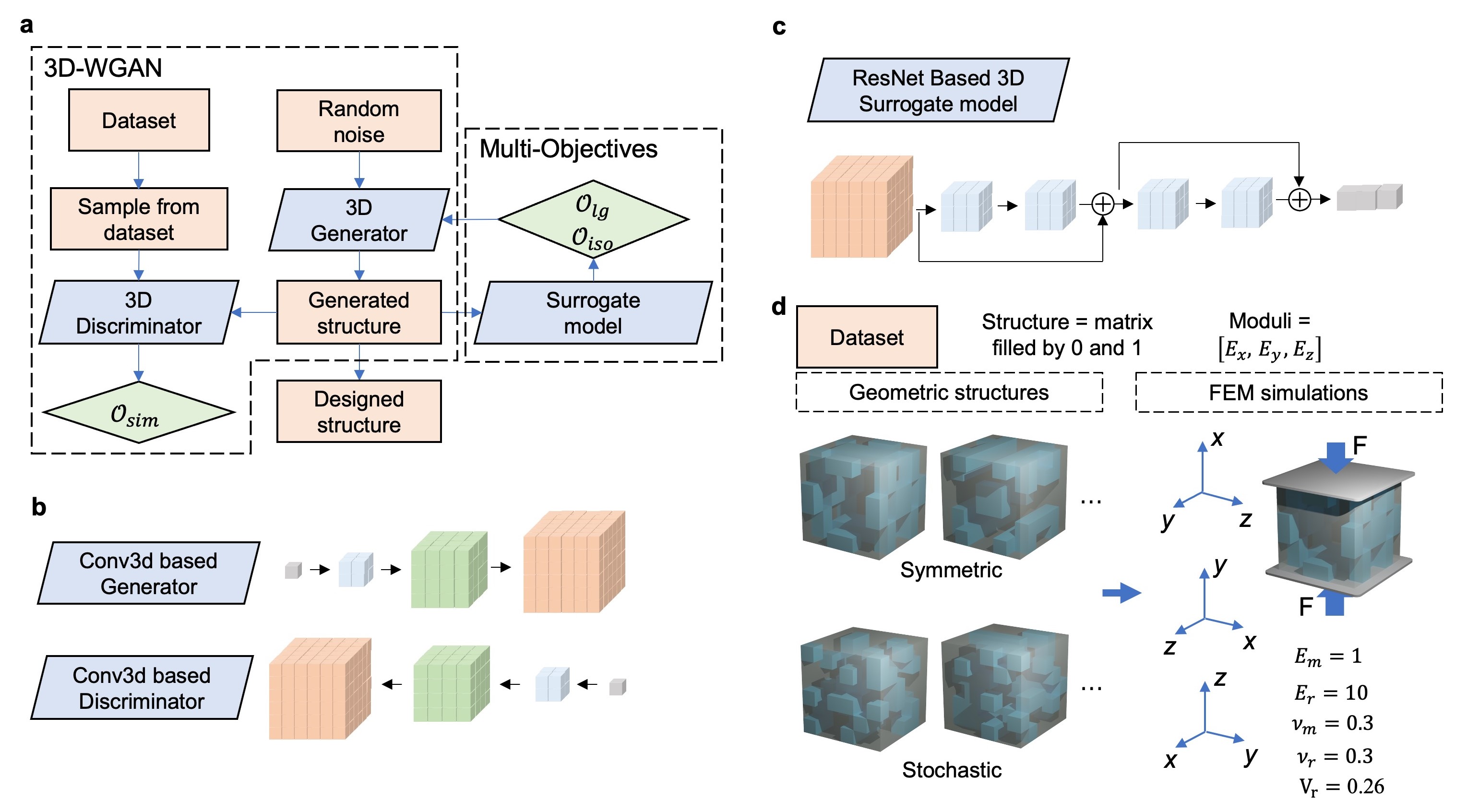}
\caption{Schematic illustration of the framework for multi-objective generative design of 3D composite materials. {\bf{a}}. Overall framework. It is mainly a GAN based model whose generator and discriminator are build by 3D convolutional neural networks instead of 2D ones. $\mathcal{O}_{sim}$ is the objective of generating similar structures with the dataset. A pre-trained surrogate model is introduced to predict the moduli of output composite structures from the generator. $\mathcal{O}_{lg}$ and $\mathcal{O}_{iso}$ are the objectives introduced to control the moduli and isotropy, separately. {\bf{b}}. Dataset construction. Voxel model of 20,000 stochastic and 20,000 center-symmetric composite structures are generated in a 6$\times$6$\times$6 space. The blue voxels represent for the reinforcements and the grey voxels represent for the matrix. The volume fraction \textit{$V_{r}$}=0.26, the Young's moduli and Poisson's ratios of the matrix and reinforcements {$E_{m}$}, {$E_{r}$}, {$\nu_{m}$} and {$\nu_{r}$}, are kept the same for all the structures. Single-axial compression simulations are performed to get the elastic response of the generated composite structures in \textit{x}, \textit{y} and \textit{z} direction. The dataset is constructed of structure tensors and moduli vectors \textit{$E_{x}$}, {\it{$E_{y}$}} and {\it{$E_{z}$}} of the composite structures. {\bf{c}}. 3D Wasserstein GAN network for structure generation. 3D Transposed convolutions are used to generate new structures from random seed while 3D convolutions are used to discriminate the structures generated by traditional Wasserstein GAN loss $\mathcal{L}_{sim}^G$ and the moduli and isotropy objective losses $\mathcal{L}_{lg}$ and $\mathcal{L}_{iso}$ {\bf{d}}. Network of the surrogate model. Resnet-18 based surrogate models are used provide predicted moduli and isotropy for losses $\mathcal{L}_{lg}$ and $\mathcal{L}_{iso}$ calculation. The dataset construction by geometry generation and FEM simulations. }
\label{framework}
\end{figure}

Our 3D Multi-objective Driven Wasserstein GAN (MDWGAN) framework (Figure \ref{framework}a) mainly consists of two key components: a generative network with multiple objectives, and a pre-trained 3D surrogate model to give evaluations for the objectives. Based on the the idea of Wasserstein GAN (WGAN)\cite{bib42_Arjovsky2017} and its modified version Wasserstein GAN-Gradient Penalty (WGAN-GP)\cite{bib44_Gulrajani2017}, we developed an extended 3D-WGAN-GP network (including 3D generator and 3D discriminator) to keep the structural similarity of generated structures and structures in the dataset. Furthermore, multiple objectives are introduced and combined with the original WGAN-GP loss function to control the properties of the composites. A surrogate model is introduced to calculate the properties real time in training iterations.

%During the training process, there are two main training routes, one for 3D-based Generator and the other for 3D-based Discriminator. For the former training route, a random noise is produced and inputted to the 3D-based Generator. Then based on the random noise, the 3D-based Generator samples from a latent space. With the generated structure, the 3D-based Discriminator produces Loss $G_1$. Meanwhile, the generated structure is also sent to the pre-trained 3D Surrogate model. Based on the objectives we chose, we can calculate Loss $G_2$ or Loss $G_3$. Both Loss $G_1$ and Loss $G_2/G_3$ will be backpropogated to the 3D-based Generator. Note that in this procedure, the parameters of 3D-based Discriminator are frozen. For the latter training route, a random noise will be passed to the frozen 3D-based Generator and transformed into a structure. Also, we samples from the dataset and inputs both the generated fake structure and the real structure into the 3D-based Discriminator. Based on the output, we calculate Loss D and backpropogate it to the 3D-based Discriminator. 

\subsubsection{Objective functions}\label{mdloss}

In our model, we mainly focus on three parts of objectives: $\mathcal{O}_{sim}$, $\mathcal{O}_{lg}$ and $\mathcal{O}_{iso}$. Each objectives gives a direction for the 3D-based generator to optimize and they are combined with different weights to achieve desired performances.

$\mathcal{O}_{sim}$ denotes the objective that imitates the data from the original training set. This objective can be also regarded as a soft constraint embedded in MDWGAN. Since we aim at generating new structure designs, we do not intended to add hard constraints to limit the volume fraction of strong phase $\it{V_{r}}$ of the composite, the geometrical symmetry, or the pattern of constituent material construction. Instead, we rely on a 3D, GAN-like network. That is, the 3D discriminator will restrain the behaviour of the 3D generator by distinguishing the structures produced by 3D-based generator from the training data, based on their similarity. Note that this is a probabilistic prediction of reinforcement phase locations given by not only one parameter such as the fraction of strong phase, but also partially inherits geometrical and structural features of the training data. 

$\mathcal{O}_{lg}$ describes the objective that the model has a tendency to generate structures with higher average moduli. % Note that this tendency will not guarantee certainty on moduli optimization, but can increase the performance at an expectation degree.
$\mathcal{O}_{iso}$ denotes the isometric moduli objective, which means that we encourage the 3D-based generator to generate structures with the same moduli in all directions. Similar to $\mathcal{O}_{lg}$, it offers a tendency to the optimization of the 3D-based Generator.

According to these objectives, we embodied those goals by the training losses. Those losses will gradually fine tune the 3D-based Generator  during the training process.

For objective $\mathcal{O}_{sim}$, we need to construct two parts of loss $\mathcal{L}_{sim}^D$ and $\mathcal{L}_{sim}^G$ denoting the loss for 3D-based discriminator and 3D-based generator, respectively. We follow the discriminator loss $\mathcal{L}_{sim}^D$ from WGAN-GP\cite{bib44_Gulrajani2017} for our 3D-based discriminator, i.e.,
\begin{equation}\label{lossd}
    \mathcal{L}_{sim}^D=D_{w}(\Tilde{x})-D_{w}(x)+\lambda(\parallel\triangledown_{\Tilde{x}}D_{w}(x)\parallel_2-1)^2,
\end{equation}
where $D_w$ is the 3D-based discriminator, $\Tilde{x}$ is the fake sample generated by 3D-based generator $G$, $x$ is the real sample from our dataset, and $\lambda$ is the gradient panalty coefficient.

Again, we take the generator loss $\mathcal{L}_{sim}^G$ from WGAN-GP\cite{bib44_Gulrajani2017}, i.e., 
\begin{equation}\label{LG1}
    \mathcal{L}_{sim}^G=-D_{w}(\Tilde{x}),
\end{equation}

As mentioned in Figure \ref{framework}, we have designed multiple additional losses for different objectives. We construct the loss $\mathcal{L}_{lg}$ for objective $\mathcal{O}_{lg}$ as follows:
\begin{equation}\label{LG2}
    \mathcal{L}_{lg}=w_{lg}\cdot\frac{1}{B}\sum_{i=1}^B\parallel E_{goal}-S(\Tilde{x_i})\parallel^2_2,
\end{equation}
where $w_{lg}$ is a hyperparameter that denotes the weight of loss $\mathcal{L}_{lg}$, $B$ is the batch size, $E_{goal}$ is the modulus goal we set. $S$ is the pretrained surrogate model and $\Tilde{x_i}$ is the $i$-th sample in the batch. Thus, S($\Tilde{x_i}$) is the predicted modulus of a generated sample.

Also, we construct the loss $\mathcal{L}_{iso}$ for isotropy objective $\mathcal{O}_{iso}$, which can be represented as follows:
\begin{equation}\label{LG3}
    \mathcal{L}_{iso}=w_{iso}\cdot\frac{1}{B}\sum_{i=1}^B\sum_{j,k\in\{X,Y,Z\},j\neq k} (S_j(\Tilde{x_i})-S_k(\Tilde{x_i}))^2,
\end{equation}
where $w_{iso}$ is a hyperparameter that denotes the weight of loss $\mathcal{L}_{iso}$ and $S_{X},S_{Y},S_{Z}$ denotes the calculated moduli of $X,Y,Z$ direction.

\subsubsection{3D surrogate model}\label{surr}
As introduced previously, our MDWGAN model includes a special component - 3D Surrogate Model (Fig \ref{framework}). Our 3D Surrogate Model is based on ResNet-18\cite{bib43_He2015}. Different from its original version, we switch its 2D convolutional networks to 3D convolutional networks and modify its pooling layers. We pre-trained the model to high accuracy before incorporating it to our framework (See method). Our pre-trained surrogate model can estimate the moduli of a given structure more efficiently compared to FEM (see Figure S), which allows us to significantly reduce our training time.

\iffalse
\begin{figure}[htbp]
	%\addtocounter{figure}{-1} 
	\subfigure{
		\begin{minipage}[t]{0.305\linewidth}%%%%%%%%%note2
			\centering
			\includegraphics[width=0.99\linewidth]{figures/Surrogate_result_plot_1.png}
		\end{minipage}
	}
	\subfigure{
		\begin{minipage}[t]{0.305\linewidth}%%%%%%%%%note2
			\centering
			\includegraphics[width=0.99\linewidth]{figures/Surrogate_result_plot_2.png}
		\end{minipage}
	}
	\subfigure{
		\begin{minipage}[t]{0.305\linewidth}%%%%%%%%%note2
			\centering
			\includegraphics[width=0.99\linewidth]{figures/Surrogate_result_plot_3.png}
		\end{minipage}
	}
	\caption{Comparison between real moduli and moduli calculated by surrogate model on x, y and z axes.}
	\label{surrfig}
\end{figure}
\fi

\begin{figure}[htbp]
	\centering
	\includegraphics[width=0.99\linewidth]{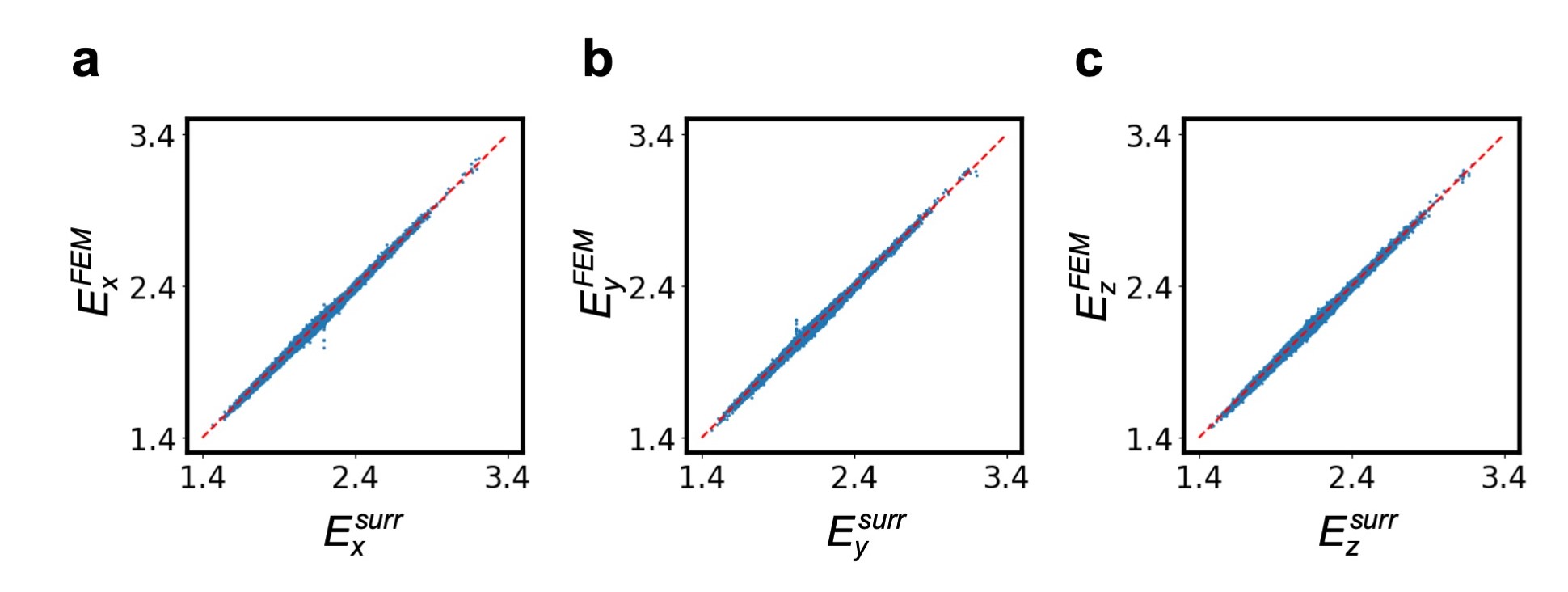}
	\caption{Comparison between real moduli and moduli calculated by surrogate model on x, y and z axes.}
	\label{surrfig}
\end{figure}
Figure \ref{surrfig} gives us a view of calculated moduli versus real moduli. Based on our calculation, the variance of the real moduli is $3.63\times 10^{-2}$, while the mean square error (MSE) between the moduli from our surrogate model and the real moduli is $2.02\times 10^{-4}$.  This result enables us to use the pre-trained 3D surrogate model to approximate the moduli of generated instance and accelerate the training process.

\subsection{Database construction}\label{database}
To illustrate our framework, we generated multiple two-phase composite structures represented by representative volume elements (RVEs) filled with cubic voxels. Reinforcement volume fractions is set as $V_{r} = 0.26$. The Young's moduli and Poisson's ratio of the matrix and reinforcements {$E_{m}$}, {$E_{r}$}, {$\nu_{m}$} and {$\nu_{r}$} are kept the same across all the structures. The Young's moduli of the composite \textit{$E_{x}$}, \textit{$E_{y}$} and \textit{$E_{z}$} are obtained (see Figure \ref{framework}b and Methods). To test our framework's capability of quality generation with few feature and parameters, only the raw structure matrix representing the voxel model of the composites and their Young's moduli vectors are included, with no other extracted features, hyperparameters or constraints such as reinforcement volume fractions $V_{r}$ or parameter of the structural symmetry and isotropy to support the model. To illustrate that our framework could generate composite structures with similar structural features, structures with both stochastic reinforcement distribution and central symmetric distributions (Figure \ref{framework}b) are included. (As there are 230 crystallographic symmetry groups in 3D spaces, to exhaustively cover all the possible symmetries in our dataset is hard and not referential in 
practical research or industrial applications.) Without the consideration of symmetry, their are $2^{n^{3}}$ possible structures in an $\textit{n$\times$n$\times$n}$ cubic voxel RVE. We select $\textit{n}$ = 6 in this work for easy distinguishable geometrical and structural features such as symmetry in generated structures.

%\begin{figure}[htbp]
%\centering
%\includegraphics[scale=0.4]{figures/Figure_data_gen.png}
%\caption{The dataset construction by geometry generation and FEM simulations. {\bf{a}}. Voxel model of 20,000 stochastic and 20,000 center-symmetric composite structures are generated in a 6X6X6 space. The blue voxels represents for the reinforcements and the grey voxels represents for the matrix. The volume fraction \textit{$V_{f}$}=0.26 is kept the same across all the structures. {\bf{b}}. Single-axial compression simulations are performed to get the elastic response of the generated composite structures in \textit{x}, \textit{y} and \textit{z} direction. The Young's moduli and Poisson's ratio of the reinforcements and the matrix are shown in the figure. {\bf{c}}. The dataset is constructed of structure tensors and moduli vectors. There are zeros representing the matrix and ones representing the reinforcements in the 6X6X6 structure tensor, while the moduli vector is a 1x3 vector filled with the Young's moduli \textit{$Ec_{xx}$}, {\it{$Ec_{yy}$}} and {\it{$Ec_{zz}$}} of the composite structures.}
%\label{data_gen}
%\end{figure} 

\subsection{Composite structures generation}\label{geometry} % Topology information?
\begin{figure}[htbp]
\centering
\includegraphics[width=0.8\linewidth]{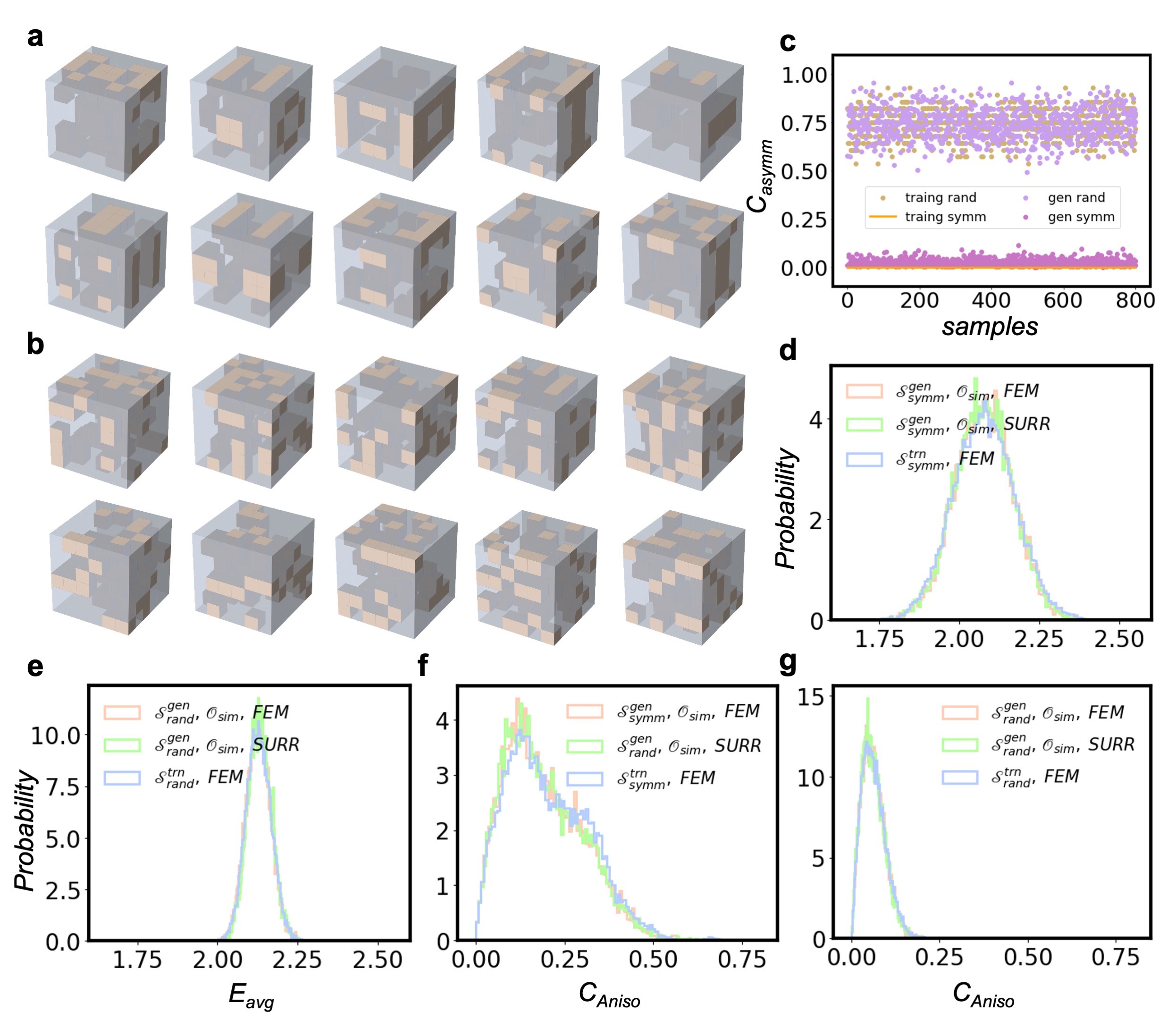}
\caption{The structures generated by MDWGAN without any additional objectives and restrictions. {\bf{a}}. Generated random and symmetric composite structures. The light orange voxels represents for the reinforcements and the light blue voxels represents for the matrix.  {\bf{b}}. MDWGAN produces composite structures with volume fractions very close to \textit{$V_{r}$}=0.26 set for the structures in training dataset. {\bf{c}}. Symmetry evaluations of the structure. {$C_{asymm}$} the fully central symmetric structures with {$C_{asymm}=0$}, while the fully stochastic structures with {$C_{asymm}=1$}. {\bf{d}}. The moduli \textit{$E_{x}$}, {\it{$E_{y}$}} and {\it{$E_{z}$}} of the generated composite structures calculated by surrogate model and FEM.}
\label{generated_pure_gan}
\end{figure}

To capture the structural features of the composites, our MDWGAN model with only $\mathcal{O}_{sim}$ but no additional physics based objectives was first trained with random and symmetric dataset. The generated structures by trained model inference are shown in \ref{generated_pure_gan}. It can be seen that the MDWGAN model not only successfully produced structures with the reinforcement volume fraction around {$V_{r}$}=0.26, but also captured the geometrical symmetry information of the structures in our dataset (Figure \ref{generated_pure_gan}a and b). Coefficient of asymmetry {\it{$C_{asymm}$}} is introduced to evaluate the symmetry of the structure . We regard the fully central symmetric structures with {\it{$C_{asymm}=0$}}, while the fully stochastic structures with {\it{$C_{asymm}=1$}} (See Method). The results show a clear difference between symmetric generations (less than 0.1, purple dots in Figure \ref{generated_pure_gan}c) and random generations (large than 0.5, red dots in Figure \ref{generated_pure_gan}c). Thus, structural features of the designs are successfully constrained by our dataset in terms of both $V_{r}$ and $C_{asymm}$. As there are no additional objectives to control the properties involved, the distribution of the Young's moduli \textit{$E_{avg}$} (Figure\ref{generated_pure_gan}d and e) as well as the \textit{$E_{x}$}, {\it{$E_{y}$}} and {\it{$E_{z}$}} (see 
Figure S1) of both symmetric and random structures generated are similar with those of the structures in our dataset. Similarly, the {\it{$C_{asymm}$}} of the composite structures generated also follow the same distribution patterns with the training datasets (Figure\ref{generated_pure_gan}f and g).

\subsection{Multi-objective composite structures design}\label{Multi-Objective}
\begin{figure}[htbp]
\centering
\includegraphics[width=0.8\linewidth]{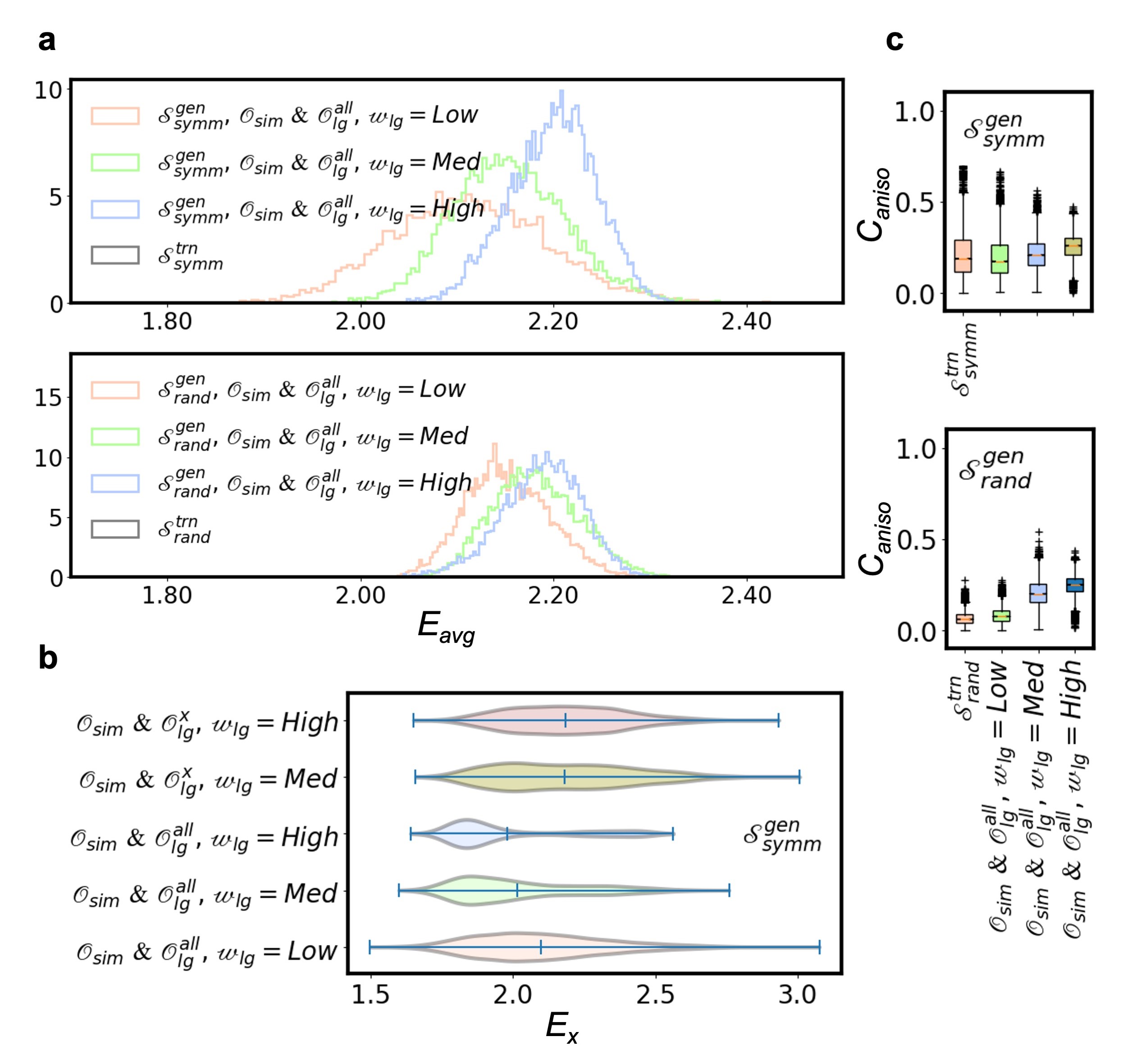}
\caption{The performance of model with $\mathcal{O}_{sim}+\mathcal{O}_{lg}$. $\bf{a}$. Distributions of average moduli for generated samples and training dataset. $\mathcal{S}_{symm}^{gen}$ denotes the moduli of generated samples of which the model has been trained on dataset with symmetrical samples. $\mathcal{S}_{symm}^{trn}$ denotes the moduli of the symmetrical samples in the training dataset. $\bf{b}$. Display of moduli on direction X for different generated samples. $\mathcal{O}_{lg}^x$ means that enforcement for moduli on direction X is applied during the training. $\bf{c}$. Box plot of anisotropic parameter $C_{aniso}$ for different generated samples. $C_{aniso}$ describes the anisotropic level for moduli of generated samples.}
\label{multi-obj-large}
\end{figure}

Based on the aforementioned $\mathcal{O}_{sim}$, $\mathcal{O}_{lg}$ and $\mathcal{O}_{iso}$, we introduced multiple objectives based on the 3D WGAN model to control the average Young's moduli and the isotropy of the composite. Firstly, combining $\mathcal{O}_{sim}$ and $\mathcal{O}_{lg}$ by low, medium and high weights (see Methods), the stiffness of the structures are tuned while keeping similar symmetric and stochastic structural features. It can be seen from Figure \ref{multi-obj-large}a that the peak of distribution of the Young's moduli moves from left to right with the increase of the weight on {\it{$\mathcal{O}_{lg}$}}. As the design space of central symmetric structures for illustration are relatively limited (approximately $1.2\times10^{5}$), the structures with $\it{E}_{avg}$ close to the upper limit are already in the training dataset. Thus, the distributions of the structures generated under low, medium and high weight of $\mathcal{O}^{avg}_{lg}$ narrows to the right side, with no significant increase in their largest value. On the contrary, the $\it{E}_{avg}$ distributions of the structures generated overall shift to the right, with increased largest values, decreased smallest values and not largely changed width, because the feasible design region of stochastic structures are much larger (approximately $3.7\times10^{51}$). We also found that besides the $\it{E}_{avg}$, $\it{E}_{x}$, $\it{E}_{y}$ and $\it{E}_{z}$ are deviated from their distribution pattern of the training dataset as the bottom three violin plots of Figure \ref{multi-obj-large}b shows. It can be found in the generated structures of some trained models that the distributions of their $\it{E}_{x}$, $\it{E}_{y}$ and $\it{E}_{z}$ split apart, and the highest peak of $\it{E}_{x}$ decreases while those of the $\it{E}_{y}$ and $\it{E}_{z}$ increase (see also Figure S2). It could partly be attributed to the pattern differentiation of the generated structures. The plane-like reinforcement distributions and three dimensional symmetric distributions of the composite both occurs in the generated structures (see generated structures with largest moduli in Figure S3 - S8). It also can be seen that the average value of $\it{E}_{x}$ decreases, while those of $\it{E}_{y}$ and $\it{E}_{z}$ increase sightly. Those might be caused by perturbations in training process. Obviously, by adding $\mathcal{O}^{x}_{lg}$ objective, $\it{E}_{x}$ could be enhanced (top two violin plots in Figure \ref{multi-obj-large}b). Although the generated structures achieve a $\it{E}_{avg}$ enhancement, the volume fraction \textit{$V_{r}$} and isotropic feature \textit{$C_{aniso}$} of the generated structures doesn't change much (Figure \ref{multi-obj-large}c). 

\begin{figure}[htbp]
\centering
\includegraphics[width=1\linewidth]{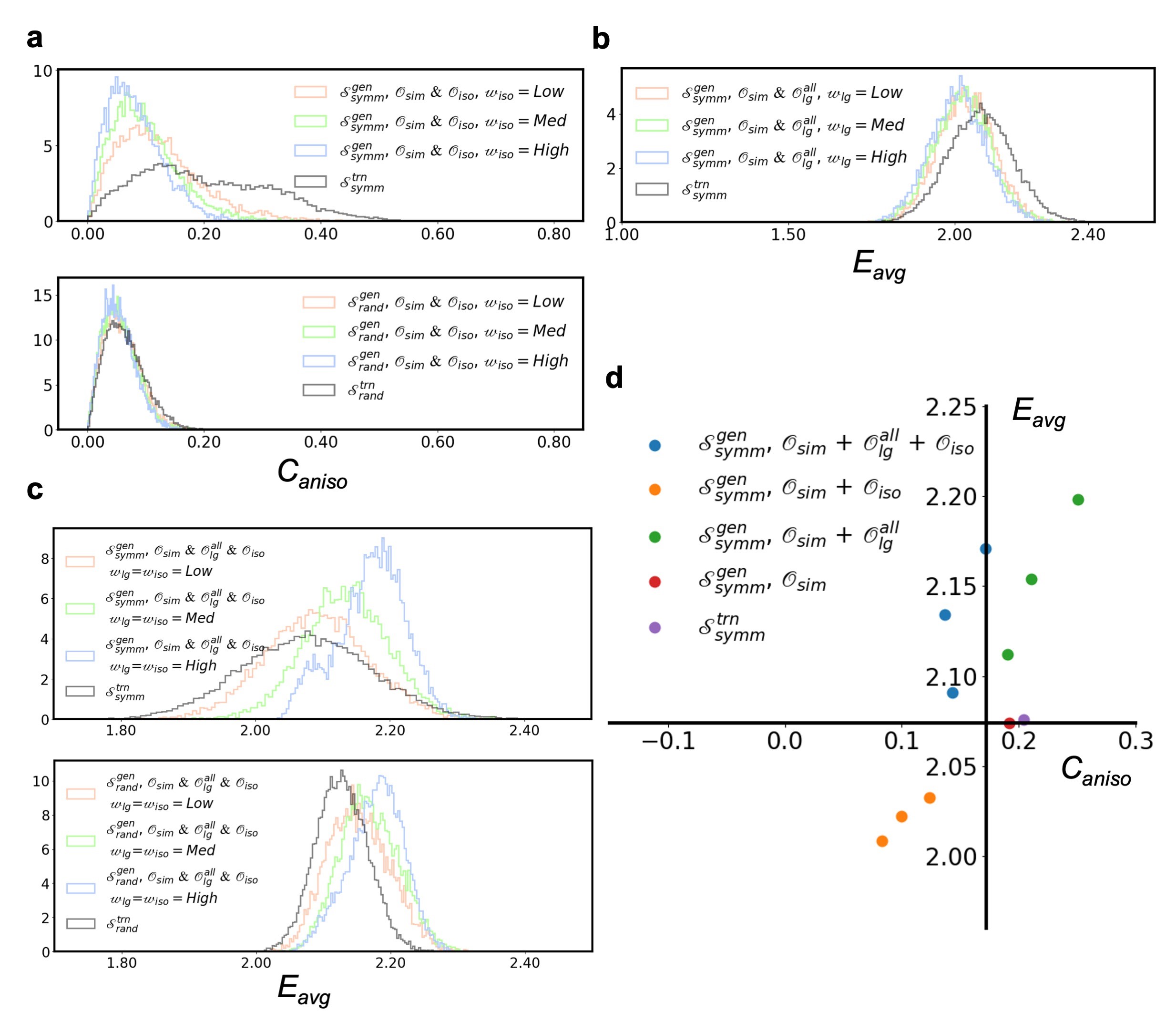}
\caption{The performance of model with $\mathcal{O}_{sim}+\mathcal{O}_{iso}$. $\bf{a}$. Distributions of anisotropic parameter $C_{aniso}$ for generated samples and training dataset. $\bf{b}$. Distributions of average moduli for generated samples and training dataset.}
\label{multi-obj-iso}
\end{figure}

The isotropy of the composite is improved by adding {\it$\mathcal{O}_{iso}$} to the {\it$\mathcal{O}_{sim}$} objective (Figure \ref{multi-obj-iso}a). From the statistical data of our generated symmetric structures, the $C_{aniso}$ we defined (see Methods) can be remarkably reduced by rising the weight of $\mathcal{O}_{iso}$, while this effect is not noticeable on stochastic structure generations as stochastic structures are almost isotropic themselves (see Figure S9). 

We found that the $\it{E}_{avg}$ of the structures generated under $\mathcal{O}_{sim}$+$\mathcal{O}_{iso}$ are marginally decreased (Figure \ref{multi-obj-iso}b). Therefore, $\mathcal{O}_{sim}+\mathcal{O}_{lg}+\mathcal{O}_{iso} $ is used to reconcile this conflict-like effect (Figure \ref{multi-obj-iso}d)and obtain composite structures with simultaneously higher Young's moduli and isotropy (Figure \ref{multi-obj-iso}c). 

\iffalse
\begin{figure}[htbp]
\centering
\includegraphics[width=0.65\linewidth]{figures/fig6.jpg}
\caption{The performance of model with $\mathcal{O}_{sim}+\mathcal{O}_{lg}+\mathcal{O}_{iso}$. $\bf{a}$. Distributions of average moduli for generated samples and different training datasets. $\bf{b}$. Average moduli and anisotropic parameter of the generated samples with different training objectives. Note that different points in the same color means training with the same objective but with different weights.}
\label{multi-obj-total}
\end{figure} 
\fi

%\input{conclusion}

\section{Discussion}\label{sec12}
\begin{figure}[htbp]
    \centering
    \includegraphics[width=0.9\linewidth]{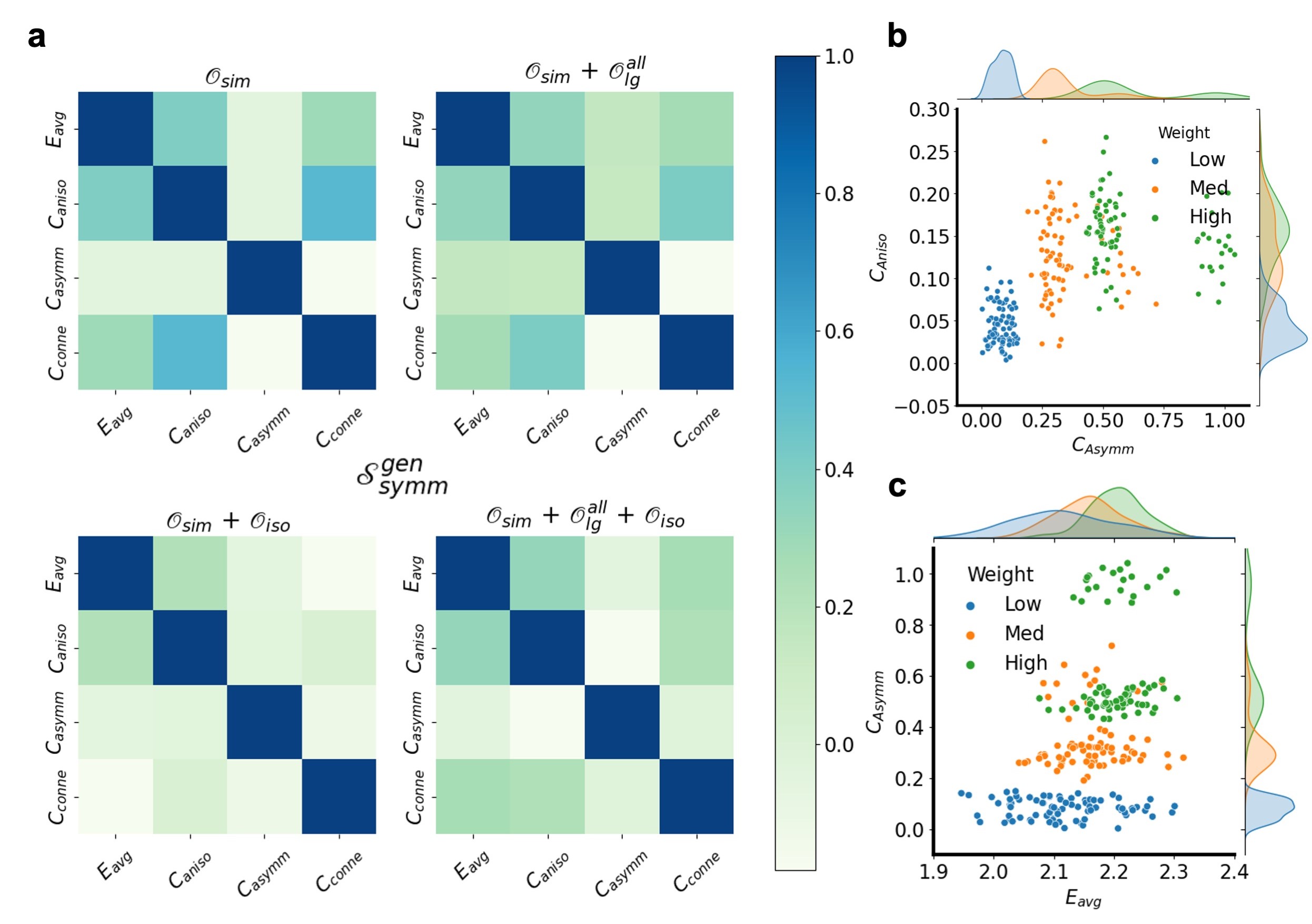}
    \caption{Correlation between the features of generated symmetric structures. \textbf{a}. Heatmap of correlation between $\it{E}_{avg}$, $\it{C}_{Asymm}$, $\it{C}_{Aniso}$ and $\it{C}_{conne}$. The shade degree of the color in each small square denotes the correlation between different parameters. Brunet squares means two parameters has high probability of correlation, both positive related or negative related. \textbf{b}. Scatter map of the correlation between $\it{C}_{Asymm}$ and $\it{C}_{Aniso}$. \textbf{c}. Scatter map of the correlation between $\it{E}_{avg}$ and $\it{C}_{Asymm}$.}
    \label{fig6}
\end{figure}

It is worth to check the correlations between $\it{E}_{avg}$, $\it{C}_{Asymm}$, $\it{C}_{Aniso}$ of all the composite structures generated. As another common descriptor for composite structures, the connectivity \cite{bib55_Lin2022, bib56_Kechagias2022, bib57_Seif2022}, of both constituent phase are calculated (See Methods) and a coefficient of connectivity $\it{C}_{Conne}$ is set as the largest number of connected components (voxels). According to the correlation heat map (Figure \ref{fig6}a), there are no convincing correlations between any of the two parameters on both symmetric and random (see Figure S10) composite structures generated. However, investigating the correlation scatter map of $\it{E}_{avg}$ vs $\it{C}_{Asymm}$ and $\it{E}_{avg}$ vs $\it{C}_{Aniso}$, we could find that a higher weight of $\mathcal{O}^{all}_{lg}$ statistically stimulates the generation of more asymmetric and anisotropic structures (Figure \ref{fig6}b and c). The reason may lie on that the objective $\mathcal{O}^{all}_{lg}$ affects the objective $\mathcal{O}_{sim}$ during training, which make the structural features of generated structures deviates from those of the dataset. 
% In terms of connectivity, we found it higher in the structures generated with higher stiffness (Figure \ref{fig6}c). 

\begin{figure}[htbp]
    \centering
    \includegraphics[width=0.9\linewidth]{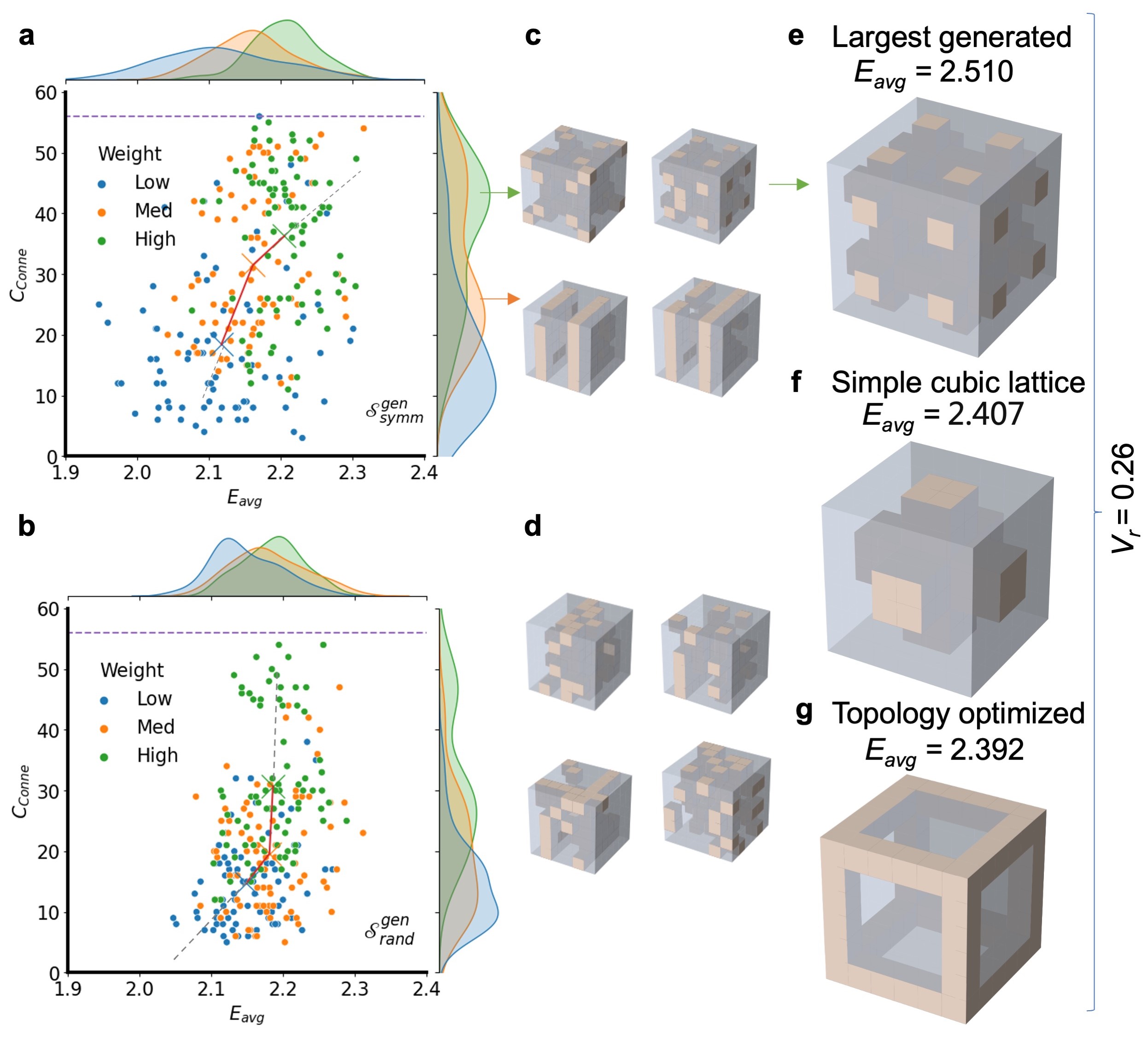}
    \caption{Analysis of the design pattern for MDWGAN. Several typical samples are sampled from the peak of distribution wave. \textbf{a}. Scatter map with k-means cluster centers of low, medium and high weight of $\mathcal{O}^{all}_{lg}$ of generated symmetric structure samples. \textbf{b}. Scatter map with k-means cluster centers of low, medium and high weight of $\mathcal{O}^{all}_{lg}$ of generated random structure samples. \textbf{c}. Typical symmetric structures generated. \textbf{d}. Typical random structures generated. \textbf{e}. Structure generated with largest $\it{E}_{avg}$. \textbf{f}. Structure and $\it{E}_{avg}$ of a common design, simple cubic lattice reinforced composite with same $\it{V}_{r}$ = 0.26. \textbf{g}. Topology optimized structure and its $\it{E}_{avg}$ with same $\it{V}_{r}$ = 0.26.}
    \label{fig7}
\end{figure}

For the $\it{E}_{avg}$ and $\it{C}_{Conne}$, we find they are, to some extent, related. We calculate the k-means cluster centers (see Method) on the scatter map between the $\it{E}_{avg}$ and $\it{C}_{Conne}$ of the 1$\%$ generated structure samples (here is 80 samples each for low, medium and high weight as we generated 8000 different structures under each weighted objective for the analysis). The k-cluster shows clear positive relates between $\it{E}_{avg}$ and $\it{C}_{Conne}$ in both symmetric and random structures generated. In the symmetric structures, two patterns of composites structures to enhance their Young's moduli are observed: to align the reinforcements in two axes (e.g., x and y) and to align the reinforcements in all three axes (x, y and z). When the weight of objective $\mathcal{O}^{all}_{lg}$ is set as medium, the majority of the structures follows the two-axes alignment pattern of the reinforcements. The distribution peak of the maximum connected volume fraction of the reinforcement is around 0.13 (1/2 of $\it{V}_{r}$). However, when the weight of objective $\mathcal{O}^{all}_{lg}$ goes to high, the three-axes alignment pattern takes the dominate position. The distribution peak of the maximum connected volume fraction of the reinforcement is close to $\it{V}_{r}$ = 0.26. This also corresponds to the two different structural paradigms shown on the right side of Figure \ref{fig7}c  (also see Figure S3 - S8).

The framework is capable of generating composite structures with similar, or even slightly larger Young's modulus than commonly regarded high stiffness structures (Figure \ref{fig7}e and f). For complex problems such as $\it{E}_{avg}$, the framework is capable of achieving the same or even better result than a robust topology optimization algorithm \cite{bib58_Sigmund2013} (Figure \ref{fig7}e and g). Knowing that the topology optimization methods work on mathematical models with clear definitions and expressions, the optimal solution is achieved by iteratively searching in the feasible region with optimal directions (see Method). Due to the complexity of the optimization model, one is likely to obtain several different optimized structures with different searching strategies. The performance of these optimized structures could be very close; thus they are regarded as the local optima in certain conditions. Therefore, each structure generated of proposed framework could also be considered as local optima in optimization space, but with large amount of samples, they are much closer to the global optima. It is also worth to mention that the FE evaluation of the modulus in topology optimization is slightly different from that in our framework in training and dataset construction. This may cause different optimization results. To systemically consider the advantages and disadvantages of our framework and topology optimization, and even integrate both to form a more effective and powerful tool for structure design should be a valuable direction for the future work.

Besides, it should be admitted that the dataset we constructed is more for framework illustration purpose. The selection or construction of a more well-developed, more informative material dataset should be of importance to demonstrate the full capability of this framework. Another future direction worth mentioning is to add active learning method into our training iterations for a more stable and mathematically complete framework.

In summary, we develop a 3D generative design framework for composite materials which consists of 3D WGAN, ResNet based surrogate, and multiple objectives of desired materials properties. It is demonstrated in this work that our MDWGAN framework successfully generates composite structures which are geometrically similar to those in our dataset (in terms of what is fixed such as size, reinforcement volume fraction $\it{V_{r}}$ and symmetry), but with tunable properties governed by multiple objectives. The properties of the composite in those objectives can be tuned by adjusting their weights. This framework could obviously be used to design single-phase architectured materials, or composites consist of more than two phases with very minor adjustment. By changing the physics based ojbectives, the control and tune of the thermal, electrical and other material properties can also be achieved. Moreover, there is great potential for this framework in designing more complicated material structures such as complex structural materials, biomaterials, and energy storage materials with the input of proper datasets and a little tailoring of the framework.

% Journal npj computational materials has no conclusion section
% \input{conclusion}

\section{Methods}\label{sec14}

\subsection{Data generation}
In this work, we represent composite structures by voxel models. Discrete number zeros and ones representing for the matrix and the reinforcements of the composites are put in the corresponding indices of 3D structure matrices as structure data points. The volume fraction of the reinforcement is set as 56/216 $\approx$ 0.26. 20000 random structures are generated by randomly locating reinforcements in the RVE. 20000 structures with central symmetric reinforcements are generated by randomly locating 7 reinforcement voxels in 3$\times$3$\times$3 RVE and mirror the structure on \textit{x}, \textit{y} and \textit{z} axes. For simplicity, the constituent materials are regarded as isotropic. The Young's moduli of the reinforcement materials $E_{r}$ are normalized by the Young's moduli of the matrix $E_{m}$, namely $E_{m}$ = 1 and $E_{r}/E_{m}$ = 10. Numerical single-axial compression test via FEM are performed three times on \textit{x}, \textit{y} and \textit{z} direction of each generated structure. The compression is controlled by displacement of each directions applied on the corresponding surfaces while the opposite ones are fixed on the same directions. The Poison's ratio used here are $\nu_{m}$ = $\nu_{r}$ = 0.3. The Young's moduli of the composite \textit{$E_{x}$}, \textit{$E_{y}$} and \textit{$E_{z}$} obtained are placed into one by three vectors. The voxel tensors and their Young's moduli vectors compose the dataset. 

\subsection{Training methodology}

Before training our key framework, we train the 3D surrogate model in advance. It follows the basic regression model training procedure, by utilizing a loss function to regular the behaviour of model. MSE loss is applied as the loss function of 3D surrogate model training.

Our training methodology of the key framework follows the idea of WGAN-GP\cite{bib44_Gulrajani2017} and implements some modifications. For every training batch, we train 3D-based generator and 3D-based discriminator in a separate way. 

For training of the 3D-based discriminator, we first initialize a random noise and pass the random noise to the frozen 3D-based generator. 3D-based generator generates a 'fake' composite structure by sampling high-dimensional noise from its latent space. Then the 3D-based discriminator outputs its judgement on the generated fake sample. Meanwhile, a real sample is also passed into the 3D-based discriminator. Loss is then composed by three parts: fake sample loss, real sample loss and gradient penalty, which is presented in Eq. \ref{lossd}.

As for the training of 3D-based Generator, its training route is more complex since we expect it to accomplish multiple objectives. Again, we sample a high-dimensional random noise and pass it to the 3D-based generator. The 3D-based generator outputs a fake sample and the frozen 3D-based discriminator judges the performance of the fake sample. Then as introduced in section \ref{mdloss}, we have different loss combinations for different objectives. For the objective of simply imitating the dataset, we just apply the loss $\mathcal{L}_{sim}^G$, which is shown in Eq. \ref{LG1}. As mentioned in section \ref{mdloss}, this loss perform as a soft constraint to limit fraction of the strong phase and the geometrical information of the structure. Meanwhile, based on the different sample data format, this loss ensures the pattern of our generated sample, like symmetric or random. If we select structure with larger moduli as our objective, we will formulate the generator loss as $\mathcal{L}_{sim}^G + \mathcal{L}_{lg}$, which corresponds to the objective $\mathcal{O}_{sim} + \mathcal{O}_{lg}$. For the objective that seeks to generate models with isometric moduli, we will formulate the generator loss as $\mathcal{L}_{sim}^G + \mathcal{L}_{iso}$. It is also practical to formulate the generator loss in the form of $\mathcal{L}_{sim}^G +\mathcal{L}_{lg} +\mathcal{L}_{iso}$, of which the performance is demonstrated in the Results section.

To make the update direction more precise, we iterate the training procedure of 3D-based discriminator multiple times but that of 3D-based once in one training batch. This strategy ensures 3D-based discriminator to keep leadership in the adversarial procedure and guide the behaviour of the 3D-based generator.

\subsection{Data processing}
In this work, $E_{avg}$ is the arithmetic mean of $E_{x}$, $E_{y}$, and $E_{z}$. $C_{aniso}$ is gauged by the standard deviation of $\it{E_{x}}$, $\it{E_{y}}$ and $\it{E_{z}}$. $C_{asymm}$ is calculated by the average difference between the structure matrix $M_{s}$ and its flipped values along $\it{x}$, $\it{y}$ and $\it{z}$ axes. Structural connectivity is performed by cc3d package \cite{bib54_cc3d} following 6-connected neighborhood method in 3D . K-means cluster centers are derived by clustering module of scikit-learn package following classic k-means algorithm \cite{bib53_scikit-learn}.

\subsection{Topological optimization}
In this work, the density-based topology optimization method is employed as a comparative example. The SIMP approach (Solid Isotropic Material Penalization) \cite{bib59_Sigmund1999} is applied to the FEM voxel model. The material properties of each element are modeled of a power-law with the so-call pseudo-density $\rho\in [0,1]$. $\rho$ = 0 or 1 represents the material is matrix or reinforcement, separately. The penalty factor p is chosen as p=3 to ensure a clear binarized 0 and 1 solution. After setting the Young’s modulus $\it{E_{avg}}$, volume fraction as objective and constraint function, the optimization problem can be solved with the OC (optimality criteria) method \cite{bib60_Bendsoe1995}. The whole optimization algorithm is realized based on a compact MATLAB code package \cite{bib61_Sigmund2001}.

\bibliography{ref}% common bib file
\bibliographystyle{unsrt}
% \input sn-bibliography.bbl

% \input sn-sample-bib.tex%

\end{document}